# CARBON NANOTUBES IN MICROELECTRONIC APPLICATIONS

FRANZ KREUPL, GEORG S. DUESBERG, ANDREW P. GRAHAM, MAIK LIEBAU,
EUGEN UNGER, ROBERT SEIDEL, WERNER PAMLER, WOLFGANG HÖNLEIN

*Infineon Technologies AG, Corporate Research,Otto-Hahn-Ring 6, 81739 Munich, Germany,
franz.kreupl@infineon.com*

Carbon nanotubes with their outstanding electrical and mechanical properties are suggested as interconnect material of the future and as switching devices, which could outperform silicon devices. In this paper we will introduce nanotubes, specify the applications, where nanotubes can contribute to the advancement of Moore's law and show our progress of nanotube process integration in a microelectronic compatible way. The growth of single individual nanotubes at lithographically defined locations on whole wafers as a key requirement for the successful implementation of nanotubes is shown. In terms of nanotube transistors we propose a vertical nanotube transistor concept which outperforms the ITRS requirements for the year 2016. The performance is mainly limited by contact resistances, but by comparison with silicon devices we show that fabricated nanotube transistors already today exceed the values for transconductance, on-resistance and drive current of silicon devices.

## 1 Introduction

Carbon nanotubes (CNTs) are a new form of carbon, discovered 12 years ago, which can be thought of as a rolled-up sheet of hexagonal ordered graphite formed to give a seamless cylinder. They can be 0.4 – 100 nm in diameter with lengths up to 1 mm. Several single-walled nanotubes (SWCNTs) can be concentrically nested inside each other, like a Russian doll, forming so-called multi-walled carbon nanotubes (MWCNTs). Due to the variety of extraordinary properties exhibited by carbon nanotubes, a large number of possible applications have been proposed [1]. In particular, the high current carrying capacity and mechanical stability of metallic nanotubes indicates applications in microelectronic interconnects [2] whereas the reasonably large band gap of narrow single-walled nanotubes suggests their use as nanoscale transistor elements [3].

When we think about alternative approaches for the fabrication of microelectronic circuits, a pre-condition for new materials is, that they have to outperform the current technology. In principle this is true for CNTs, but one of the major hurdles to overcome, is the targeted placement of a specific CNT with prescribed character, i. e. MWCNT or SWCNT, diameter and chirality. The latter determines whether a SWCNT is metallic or semiconducting. Therefore, the progress in CNT-science has shifted in recent years from a mere scientific understanding to integration issues [4,5]. Here we describe our approach to grow nanotubes on a wafer exactly where we want them to be, establishing the most advanced integration scheme for CNTs. We divide the applications in two sections, where one is devoted to the interconnect-topic, i. e. the on-chip wiring of the conventional transistors, and the other is device-related, where SWCNTs are used as switching devices.

## 2  Nanotubes in Interconnect Applications

If we look at the cross-section of a typical chip like in Fig. 1 (a), we see that nowadays chips have become "all wire". The transistors at the bottom make up only a fraction of the total chip, and already today, the speed and performance of such chips is mainly limited by the interconnects, i. e. the copper-based wiring of the transistors with different metal layers (wires) and the vertical connections between these layers, which are termed via. These vias are prone to electromigration failures as can bee seen in Fig. 1(b). The arrows mark regions, where voids have formed due to the high current densities in these structures. In 2013 the ITRS [6] predicts a current density of $3.3\ 10^6$ A/cm$^2$ in a via, a value which, to date, can only be supported by CNTs, where current densities exceeding $10^9$ A/cm$^2$ have been reported in nanotubes without heat sinks. At this ITRS technology node a MPU/ASIC half-pitch of 32 nm is predicted. On this scale, traditional interconnect schemes become problematic due to the increased wire resistances resulting from grain and surface scattering effects and the higher current densities which must be carried [7]. Sufficient heat removal from the chip is already a problem in present day computers. Due to their superb thermal conductivity, which exceeds that of diamond by a factor of two, nanotubes may also help to remove the heat more efficiently from the chip. Therefore we propose CNTs, as shown in Fig. 1(c), to realize such critical vias and contact holes.

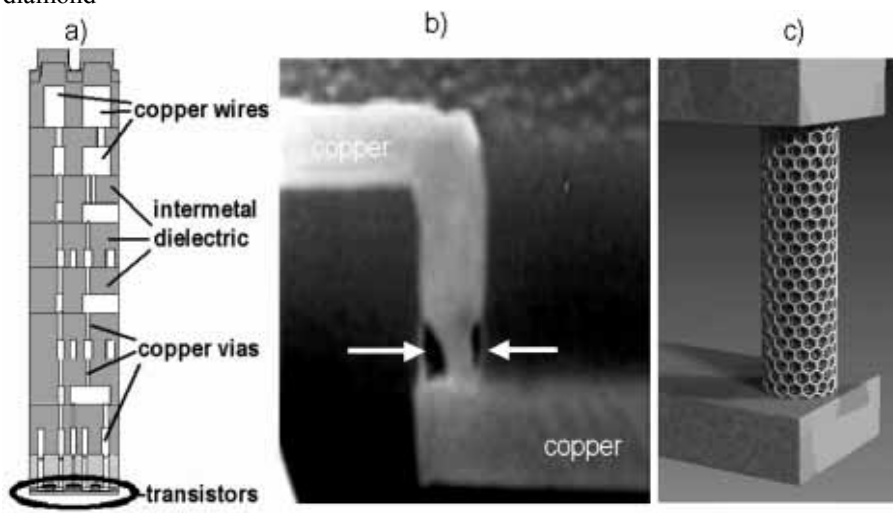

**Figure 1.** a) Cross-section through a typical chip, which consists mainly of copper-wires and –vias. b) Copper-via connecting two different metallisation levels in chip. The arrows indicate electromigration induced failures. c) Proposed CNT-via, which should withstand a 1000 times higher current density.

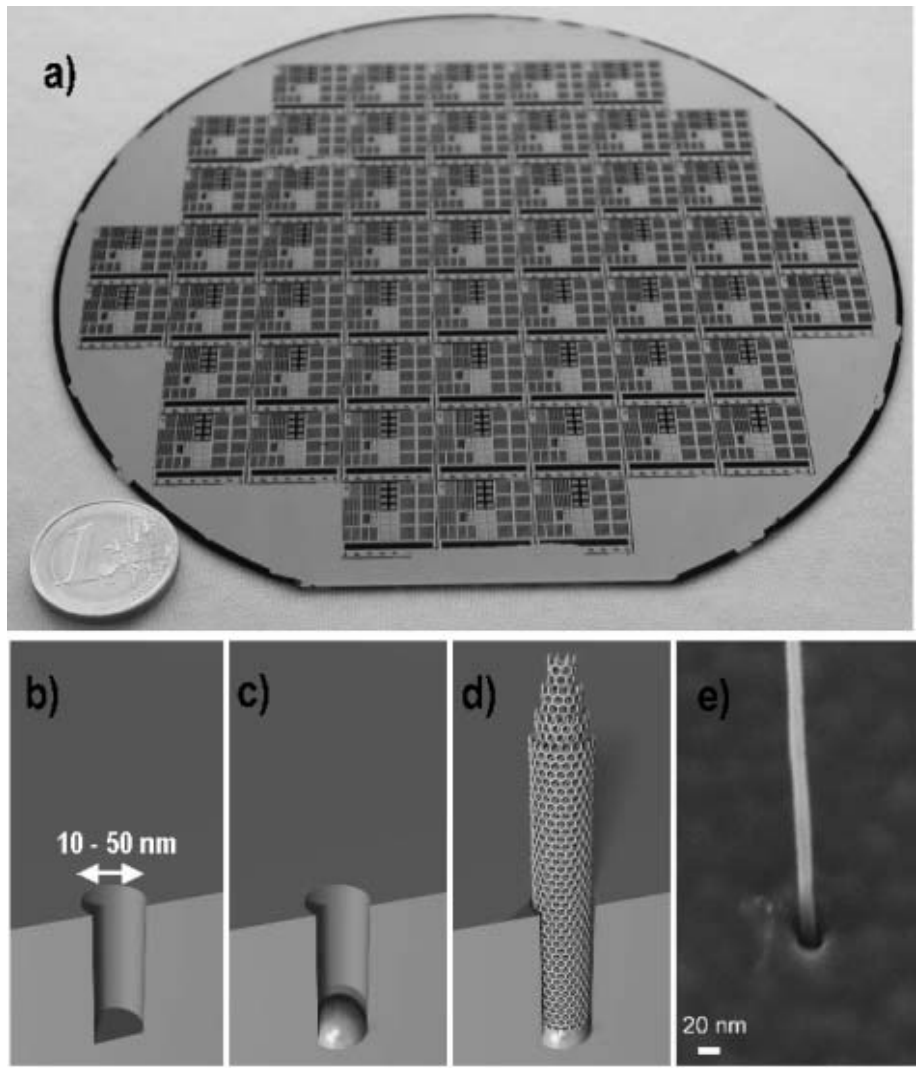

**Figure 2**. a) A 6-inch wafer with CVD-grown CNTs at lithographically defined locations. b)- c) In a 10-50 nm wide nano-hole a catalyst is deposited and MWCNTs are grown. e) A single MWCNT of 20 nm diameter protruding from a nano-via.

We have already indicated in [2], that such interconnects outclass conventional copper metallization at this reduced dimension with respect to electrical resistance and current carrying capacity.

Substantial progress has been made in the recent year by demonstrating the lithographically defined growth of CNTs on wafer-scale and the growth of individual MWCNTs in nano-vias, which have been created by conventional lithography and dry-etching methods. In Fig. 2(a) the black structures on the 6-inch wafer consists of MWCNTs, which are grown in situ using a iron-based catalyst and hydrogen-acetylene mixture as carbon source. In Fig. 2 (b)-(d), the process flow to fabricate individual CNTs at lithographically defined locations is sketched. Conventional i-line lithography in combination with a spacer-technique is used to create nano-vias with 10-60 nm diameter [8]. After the deposition of a iron-based catalyst at the bottom of the via, a single MWCNT can be grown out of the via. This can be seen in Fig. 2(e), where a 20 nm diameter MWCNT protrudes out of the nano-via. It has been observed that the diameter of the tube adjusts automatically to the diameter of the hole [8], which results in a filling factor of the via of 100%.

Challenges within this approach lie in the deposition and material of the catalyst, limited temperature budget in combination with high quality requirements for the CNTs, which normally need temperatures above 600°C to get structurally and electrically adequate results.

## 3   Carbon Nanotubes in Transistor Applications

If a semiconducting SWCNT of about 1 nm diameter is attached to two separated (metallic) contacts (source and drain), a near by third gate-electrode can modulate the conductivity of the tube by about 6 orders of magnitude at room temperature. This effect has been observed already in 1998 and has led to a kind of race in the scientific community to achieve the best performing CNT-device [3,4,5]. Although it is not yet clear, how the device actually works, the most recent work [5] can be fairly explained by the assumption of simple 1-dimensional electrostatics [9], which relates the charge in the tube by the capacitance of the tube- and gate structure and the applied gate-voltage. Based on this theory a best performance projection for CNT-transistors can be made and compared to the ITRS requirements of the year 2016. We propose a vertical, coaxially gated nanotube transistor [10], as shown in Fig. 3(a), with a 1 nm diameter tube, 10 nm gate-length and a 1 nm thick silicondioxide as the effective gate-oxide. In order to compare with ordinary silicon devices, which are always scaled to device width, we make a parallel array of this device comprising 250 CNTs per micron, as shown in Fig. 3(b). With the theory of Guo et al. [9] we can estimate the performance of the CNT-transistor and the results are tabulated and compared with the ITRS in Table 1.

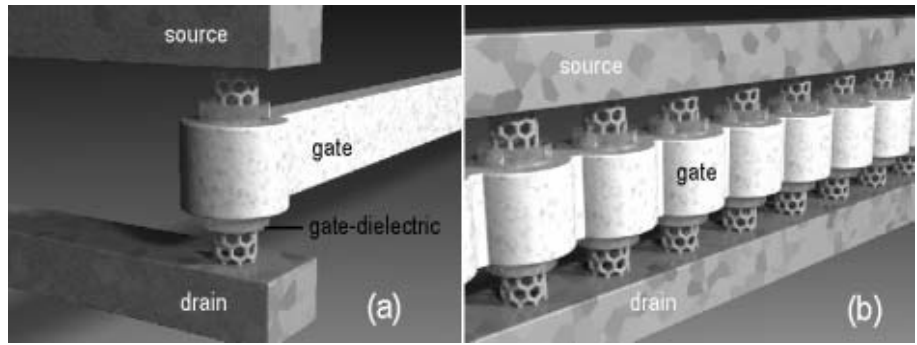

**Figure 3**. Proposed vertical coaxially gated CNT-transistor in a single (a) or parallel array (b) CNT-transistor.

It clearly can be seen that the CNT-transistor fulfills all the requirements by far. The drive current at the supply voltage of $V_{dd} = 0.4$ V is almost twice as high, the transconductance $g_m$ is almost 15 times higher, while the gate-delay t is almost half of the allowed value. The subthreshold swing S is close to the theoretical limit, the leakage current is ¼ of the allowed value and can be adjusted by the gate-material. The allowed effective equivalent gate-oxide thickness of 1 nm is well in the range of low gate leakage and manufacturability.

These promising values leave room for performance loss due to deviation from the ideal behavior. The main contribution in performance loss comes from neglecting the contact resistance, which arises between the metallic contacts and the carbon nanotube and is caused by k-vector mismatch and/or Schottky-barriers. In the following we model this resistance as linear, i. e. ohmic resistance and calculate the performance dependence on the contact resistance. The extrinsic transconductance $g_m$ can be calculated from the intrinsic transconductance $g'_m$ and the extrinsic output conductance $g_{ds}$ and is given by:

$$g_m = \frac{g'_m(1 - g_{ds}(R_s + R_d))}{(1 + g'_m R_s)}$$

**Tabel 1**. Comparison of the year 2016 ITRS requirements with the properties of the proposed vertical CNT-transistor array.

|  | Vdd Volt | drive current µA/µm | trans-conductance µS/µm | t ( Cgate* Vdd/Idd) (ps) | S mV/dec | leakage µA/µm | effective tox (nm) |
|---|---|---|---|---|---|---|---|
| ITRS Year 2016 | 0.4 | 1500 | 1000 | 0.15 | 70 | 10 | 0.4- 0.5 |
| CNT-FET | 0.4 | **2500** | **15000** | **0.08** | **65** | **2.5** | **1** |

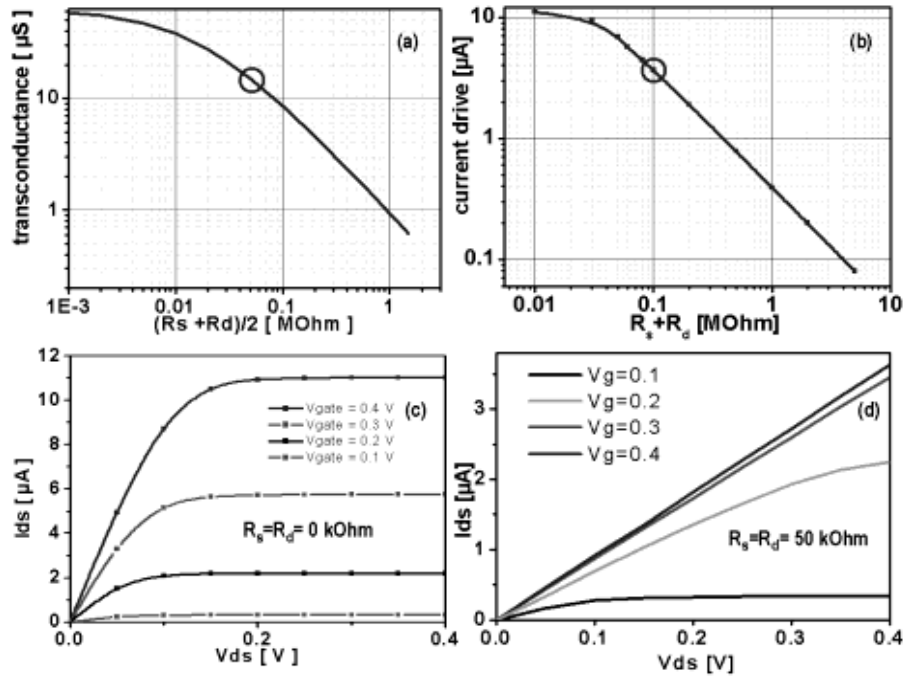

**Figure 4**. (a) The extrinsic transconductance $g_m$ as a function of symmetrical contact resistances. (b) The decrease of drive current versus contact resistance. (c) The ideal transistor characteristic compared to (d), where a contact resistance of 50 kOhm is assumed. The circles in (a) and (b) denote the respective values for the case of $R_S = R_D = 50$ kOhm.

The situation is summarized in Fig.4, where the dependence on drain- and source-contact resistances, denoted by $R_D$ and $R_S$, for the extrinsic transconductance and the current drive is shown. Experimental values for the individual contact resistances range between 30 kOhm and 2 MOhm. For an assumed resistance of 50 kOhm, the change of the ideal characteristic to that of the non-ideal, is depicted in Fig. 4(c) and (d) and the influence on the transconductance and drive current is indicated by circles in Fig. 4(a) and (b).

The proposed CNT-transistor can fulfill the ITRS requirements even with these reduced performance values. If we take into account, to scale the transistor not only by width, but by the used area, as shown in Fig. 5, we can imagine a 2-dimensional vertical array of individual nanotubes, creating a very promising device. If we compare this CNT-device with the silicon world, we have to keep in mind that the silicon device needs area for source and drain contacts and is not stackable. Whereas our proposed CNT-transistor incorporates already source and drain contacts and is stackable. So, with this concept, we can create real 3-dimensional electronics.

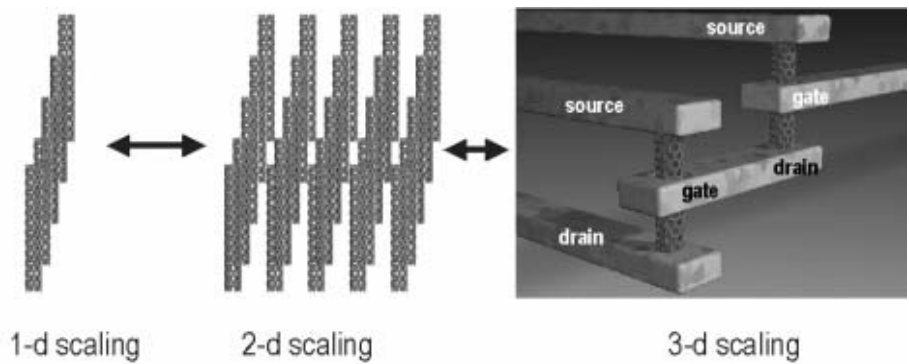

**Figure 5.** Different scaling scenarios for comparing nanotube transistors with the silicon world. The comparison should include the scaling to the used area, which favorites 2-d scaling, and the option to make real 3-d electronics.

To top off the discussion, we compare the current status of real fabricated nanotube transistors with the best performing silicon transistors in Tab. 2. After the forgoing discussion, we think that it is justified to scale the transistor properties by the device width. The CNT-FET of [11] uses electrolyte gating and can be seen as a limiting case for CNTs regarding the use of high-k dielectrics. It should be noted that we have listed only properties responsible for the static performance of the devices, as the gate-scaling is not yet as advanced as in the silicon world.

The next question to be answered for these superior CNT-transistors is what will happen with these outstanding performance values at a gate-length of 10 nm ?

**Tabel 2.** Comparison between fabricated nanotube and silicon transistors.

|  | p-CNT FET [11] 1.4 µm (1 V) Rosenblatt (2002) | p-CNT FET [5] 3 µm (1.2 V) Javey (2002) | MOSFET[12] 0.1µm (1.5V) Ghani (1999) | FinFET [13] 10 nm (1.2V) Yu (2002) | MOSFET[14] 14 nm (0.9V) Doris (2002) |
|---|---|---|---|---|---|
| drive current Ids (mA/µm) | 2.99 | 3.5 | 1.04 nFET 0.46 pFET | 0.450 nFET 0.360 pFET | 0.215 pFET |
| transconductance µS/µm | 6666 | 6000 | 1000 nFET 460 pFET | 500 nFET 450 pFET | 360 pFET |
| S mV/dec | 80 | 70 | 90 | 125 101 | 71 |
| on-resistance Ohm/µm | 360 | 342 | 1442 nFET 3260 pFET | 2653 nFET 3333 pFET | 4186 pFET |
| gate- length nm | 1400 | 2000 | 130 | 10 | 14 |
| normalized gate-oxide 1/nm | 80/1 = 80 | 25/8 = 3.12 | 4/2 = 2 | 4/1.7 = 2.35 | 4/1.2 = 3.33 |
| mobility cm$^2$/(Vs) | 1500 | 3000 | -- | -- | -- |
| Ioff(nA/µm) | --- | 1 | 3 | 10 | 100 |

## 4  Acknowledgements

We thank Zvonimir Gabric for expert technical assistance.